# Theory-guided Investigation on Magnetic Evolution of $MnPt_{5-x}Pd_xP$ and Discovery of anti-$CeCoIn_5$-type Ferromagnetic $MnPd_5P$

*Ranuri S. Dissanayaka Mudiyanselage,[1] Chang-Jong Kang,[2,3] Kaya Wei,[4], Zhixue Shu[5], Tai Kong[5], Ryan Baumbach,[4] Gabriel Kotliar,[2] Weiwei Xie[1]**

*[1]Department of Chemistry and Chemical Biology, Rutgers University, Piscataway, NJ, 08854, USA*
*[2]Department of Physics and Astronomy, Rutgers University, Piscataway, NJ, 08854, USA*
*[3]Department of Physics, Chungnam National University, Daejeon, 34134, South Korea*
*[4]National High Magnetic Field Laboratory, Tallahassee, FL, 32310, USA*
*[5]Department of Physics, University of Arizona, Tucson, AZ 85721, USA*

***Abstract***

We report the magnetic changes from canted antiferromagnetic to ferromagnetic orderings in anti-115-type $MnPt_{5-x}Pd_xP$ ($x$ = 1, 2, 2.5, 3, 4, and 5) and the discovery of a new rare-earth-free ferromagnet, $MnPd_5P$ by both theoretical prediction and experimental investigation. The family compounds were synthesized using high temperature solid state method and characterized to crystalize in the anti-$CeCoIn_5$ type with the space group *P*4/*mmm* exhibiting a two-dimensional layered structural feature. The magnetic property measurements indicate that the compounds ordered from canted A-type antiferromagnet in $MnPt_5P$ to ferromagnet above the room temperature with varying degrees of coercivity and magnetic moments in $MnPd_5P$ by reducing the spin orbital coupling. The results of the $MnPt_{5-x}Pd_xP$ have been analyzed in comparison to the other candidates of the 151 family of $Mn(Pt/Pd)_5(P/As)$ to understand the complex structure-magnetism relationships.

## Introduction

Ferromagnetic (FM) and antiferromagnetic (AFM) spintronics as well as the emerging fields of magnetic topological materials in atomically thin two-dimensional (2D) layers have attracted a great deal of experimental and theoretical interests among materials scientists, especially the magnetic anisotropy in 2D magnetism.[1,2] Such materials form a basis for most of the critical key quantum technology of future information. Most of materials are magnetic semiconductors or insulators, in which super-exchange interaction theories are widely used to explain the magnetism.[3] Very few low-dimensional itinerant magnetic metals which shows completely different electronic and magnetic properties from magnetic semiconductors have been reported and studied.[4–8] From a theoretical perspective, it is rather challenging to understand and, therefore, predict whether a structure containing transition metals will be Pauli paramagnetic, ferromagnetic, antiferromagnetic or even ferromagnetic.[9] Based on our previous experimental study, magnetically active 3*d* metals occupying voids in complex intermetallic frameworks give rise to low-dimensional structures of these magnetic metals, which may result in low-dimensional magnetic behaviors, such as layered ferromagnetic $MnPt_5As$ and antiferromagnetic $MnPt_5P$.[10,11]

The iso-valent $MnPt_5As$ and $MnPt_5P$ in the same crystal structure show completely different magnetic orderings. Both $MnPt_5As$ and $MnPt_5P$ crystalize in the $CeCoIn_5$-type tetragonal structure with the space group *P*4/*mmm*.[10,11] Different from $CeCoIn_5$, magnetic 3*d* transition metals occupy Ce site while late transition metal Pt locates on the In site. Similar to heavy-fermion CeCoIn5 system, $MnPt_5As$ and $MnPt_5P$ also show the large magnetic anisotropy along different directions.[10,11] The slight atomic radii differences between P and As completely changed the magnetic orderings in ferromagnetic $MnPt_5As$ and antiferromagnetic $MnPt_5P$. This motivated us to investigate the lattice parameters effects on magnetic orderings by theoretical simulation and predict the magnetic interactions in the low-dimensional itinerant magnets.

To accelerate our search for various new magnetic materials, we employed the magnetic ordering and total energy calculation first to determine the thermodynamic stable magnetic ordering. Using the predicted atomic radii, we successfully synthesized the family compounds anti-115-type $MnPt_{5-x}Pd_xP$ ($x$ = 1, 2, 2.5, 3, 4, and 5). Thus, the new ferromagnetic $MnPd_5P$ with the Curie-Weiss temperature as high as ~305 K was found. The experimental structural and magnetic characterizations on $MnPd_5P$ are consistent with the theoretical predictions.

## Experimental Section

**Synthesis:** High temperature solid-state pellet method was utilized to synthesize the $MnPt_{5-x}Pd_xP$ ($x$ = 1, 2, 2.5, 3, 4, and 5). Mn powder (Mangan, 99+%), Pt powder (BTC, -22 mesh, 99.99%), Pd powder (BTC, -200 mesh, 99.95%) and red P powder (BTC, -100 mesh, 99%) were mixed and ground in Mn: Pt/Pd: P = 1:5:1 atomic ratio. The mixture was pressed into a *d*-¼ pellet. The pellet was placed into an alumina crucible and sealed in an evacuated silica tube (<$10^{-5}$ Torr). The evacuated sample tube was then heated to 1050 °C at a rate of 40 °C per hour. After annealing for 2 days at 1050 °C, it was slowly cooled down to room temperature at the speed of 10 °C per hour. The polycrystalline $MnPt_{5-x}Pd_xP$ obtained was stable in air and moisture. The polycrystalline $MnPt_{5-x}Pd_xP$ was used for the structural characterization and physical property measurements.

**Phase Determination and Structural Analysis:** The phase identification of all $MnPt_{5-x}Pd_xP$ ($x$ = 1, 2, 2.5, 3, 4, and 5) polycrystalline samples were done by the Bruker D2 Phaser XE-T edition benchtop X-ray Powder Diffractometer equipped with Cu $K_\alpha$, λ=1.5405 Å. Data was collected over a range of Bragg angle, 2θ, from 5 to 90° at a rate of 0.008 °/s. LeBail fitting of Powder X-Ray Diffraction (PXRD) data was completed with the Fullprof Suite software for the initial phase identification and the lattice parameters determination.[12–14] Single crystal X-ray diffraction (SCXRD) experiments were conducted in D8 Quest Eco diffractometer with Mo radiation (($\lambda_{K\alpha}$ = 0.71073 Å) equipped with Photon II detector. Single crystals were mounted on a Kapton loop and measured with an exposure time of 10 s per frame scanning 2θ width of 0.5°. Structure refinement was performed in SHELXTL package using direct method and full matrix least-squares on $F^2$ model.[15,16] Anisotropic thermal parameter for all atoms were refined in SHELXTL. The VESTA software was used to plot the crystal structures.[17]

**Scanning Electron Microscope (SEM):** Chemical compositions of $MnPt_{5-x}Pd_xP$ ($x$ = 1, 2, 2.5, 3, 4, and 5) samples were analyzed using a high vacuum Zeiss Sigma Field Emission SEM (FESEM) with Oxford INCA PentaFETx3 Energy-Dispersive Spectroscopy (EDS) system. Spectrums were collected for 100 *s* from multiple areas of the crystals mounted on a carbon tape with an accelerating voltage of 20 *k*eV.

**Physical Property Measurements:** Magnetic properties of the polycrystalline $MnPt_{5-x}Pd_xP$ ($x$ = 1, 2, 2.5, 3, 4, and 5) was measured in Quantum Design PPMS Dynacool (QD-PPMS) at National High Magnetic Field Laboratory over a temperature range of 1.8 to 400 K with the applied field of 1000 Oe. Additionally, magnetic measurements of Pt doped compounds were carried out in vibrating sample magnetometer (VSM) in Quantum Design PPMS system over a temperature range of 1.8 – 600 K with the applied field of 1000 Oe. Magnetic susceptibility is defined as $\chi$ = M/H where M is the magnetization in units of emu and H is the applied field. The field dependent magnetization measurements were carried out at different temperatures of 2 K, 100 K, 200 K, 300 K, 325 K and 350 K up to 90 kOe.

**Theoretical Calculations:** The all-electron full-potential linearized augmented plane-wave (FLAPW) method implemented in WIEN2k[18] was adopted to calculate the electronic structure. Experimental crystal structure was selected for calculations. Generalized gradient approximation (GGA) of Perdew-Burke-Ernzerhof (PBE)[19] was chosen for the exchange-correlation functional. A 20 × 20 × 11 *k*-mesh was used for the Brillouin zone integration. The spin-orbit coupling (SOC) was included in the second-variational scheme. To consider the strong correlation effect in Mn 3d orbital, GGA+U was adopted within fully localized limit [20,21] The effective on-site Coulomb interaction parameter $U_{eff}$ = U – J = 5 eV was used, where the Coulomb parameter was turned out to be valid in previous DFT calculations on another correlated Mn compounds.[22]

## Results and Discussion

**Total Energy Analysis of Magnetic Ground States of $MnPt_5P$ and $MnPt_5As$:** In the search of new magnetic materials in the anti-115 family, total energy analysis was adopted to determine the stable magnetic ordering. Moreover, to further understand structure-magnetism relationships and the nature of ground state properties of the new family in anti-$CeCoIn_5$ structure type, we analyzed total energy differences between the A-type antiferromagnetic (AFM1) ordering and ferromagnetic (FM) orderings ($\Delta E_{AFM1}$-$E_{FM}$) in $MnPt_5As$ and $MnPt_5P$. The $\Delta E_{AFM1}$-$E_{FM}$ was calculated as a function of the lattice parameter *c*/*a* ratio with keeping the lattice parameter *a* constant for each compound. It should be noted that according to the experimental results in the family, the effect on lattice parameter *a* is negligible when X is changed in the $MnT_5X$ (X = P, As) system due to the 2D layered nature of these materials while the effect on *c* axis is non-negligible. The total energy differences computed from the first-principle calculations with spin-orbit coupling (SOC) and the effective on-site Coulomb parameter U of U = 0 eV and U = 5 eV on $MnPt_5P$ and $MnPt_5As$ are shown in **Figure 1.**

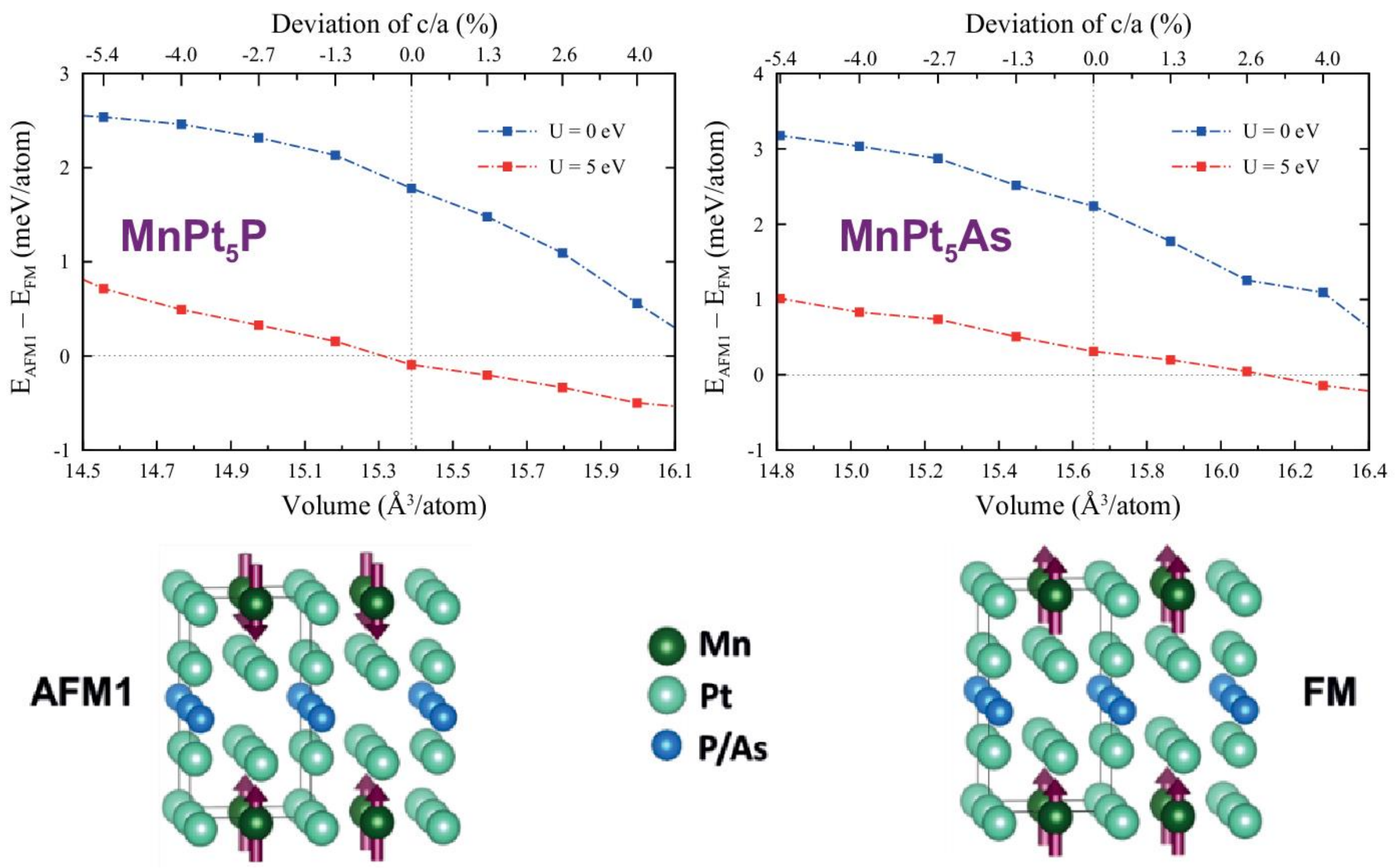

**Figure 1.** Results from GGA+SOC+U calculation by changing *c/a* ratio with a fixed lattice constant *a* for ***(a)*** $MnPt_5P$ and ***(b)*** $MnPt_5As$ with the illustration of spin orientation in ferromagnetic (FM) and the A-type antferromagnetic (AFM1) ordering.

The computational results show the magnetic ground state for both compounds. The electronic correlation affects their magnetic ground state as demonstrated in the U = 5 eV calculations (see Figure1). In the case of $MnPt_5As$, FM ordering is preferred regardless of the strength of the correlation at $\Delta c/a$ = 0.0, while the a-type AFM ordering is favorable in the presence of the correlation (U = 5 eV) above $\Delta c/a$ ~ 2.6%. On the other hand, $MnPt_5P$ shows the a-type AFM (AFM1) ordering at $\Delta c/a$ = 0.0 in the presence of the correlation (U = 5 eV), but it is close to a border of the AFM1 and FM phases. Interestingly, a new ground-state with FM ordering emerges at $\Delta c/a$ ~ -0.5%. With these theoretical predictions in mind, we aim to replace Pt with slightly smaller Pd in anti-115 system to explore new itinerant ferromagnetic material in our experiments. Initial structural and magnetic property measurements on $MnPt_{5-x}Pd_xP$ ($x$ = 1, 2, 2.5, 3, 4, and 5) confirm the magnetic changes from antiferromagnetism to ferromagnetism. Furthermore, the new anti-115 $MnPd_5P$ hosts the FM state at room temperature

**Single Crystal Structure Determination and Phase Analysis:** The variants of anti-115 $MnPt_{5-x}Pd_xP$ ($x$ = 1, 2, 2.5, 3, 4, and 5) samples were successfully synthesized using solid-state pellet reaction and structurally characterized utilizing SCXRD measurements. The resulting crystallographic data including atomic coordinates, site occupancies and equivalent isotropic thermal displacement parameters of $MnPd_5P$ is reported in the **Table 1 and 2** while crystallographic information on $MnPt_{5-x}Pd_xP$ ($x$ = 1, 2, 2.5, 3, and 4) compounds are given in the **Table S1 and S2.** The results indicate that $MnPt_{5-x}Pd_xP$ ($x$ = 1, 2, 2.5, 3, 4, and 5) crystalize in a tetragonal unit cell with the space group of *P4/mmm* the same as the previously reported $MnPt_5P$ and $MnPt_5As$ compounds in anti-$CeCoIn_5$-type structure. The refined lattice parameters for $MnPd_5P$ are $a$=3.899 (2) Å and $c$=6.867 (4). As shown in the **Figure 2*a***, $MnPd_5P$ adopts a layered structure with alternating layers of $Mn@Pd_{12}$ face sharing polyhedral and P layers.

**Table 1.** Single crystal structure refinement for $MnPd_5P$ at 300(2) K. (Standard deviation is indicated by the values in parentheses)

| Refined Formula | $MnPd_5P$ |
|---|---|
| F.W. (g/mol) | 617.91 |
| Space group; Z | *P* 4/*mmm*; 1 |
| $a$(Å) | 3.899 (2) |
| $c$(Å) | 6.867 (4) |
| V (Å$^3$) | 104.42 (9) |
| θ range (º) | 2.966-34.770 |
| No. reflections; $R_{int}$ | 578; 0.0609 |
| No. independent reflections | 170 |
| No. parameters | 12 |
| $R_1$: $\omega R_2$ ($I$>2δ($I$)) | 0.0509; 0.1204 |
| Goodness of fit | 1.282 |
| Diffraction peak and hole (e$^-$/ Å$^3$) | 2.656; -1.863 |

**Table 2.** Atomic coordinates and equivalent isotropic displacement parameters of $MnPd_5P$ at 300(2) K. ($U_{eq}$ is defined as one-third of the trace of the orthogonalized $U_{ij}$ tensor (Å$^2$)).

| Atom | Wyckoff. | Occ. | *x* | *y* | *z* | $U_{eq}$ |
|---|---|---|---|---|---|---|
| Pd1 | 4*i* | 1 | 0 | ½ | 0.2948 (1) | 0.015(1) |
| Pd2 | 1*a* | 1 | 0 | 0 | 0 | 0.012(2) |
| Mn3 | 1*c* | 1 | ½ | ½ | 0 | 0.022(1) |
| P4 | 1*b* | 1 | 0 | 0 | ½ | 0.016(2) |

**Table 3.** The comparison of structural – property features of $MnPd_5P$ with $MnPt_5P$ and $MnPt_5As$.

| | $MnPd_5P$ | $MnPt_5P$ | $MnPt_5As$ |
|---|---|---|---|
| ***c/a* ratio** | 1.761 | 1.779 | 1.816 |
| **Shortest Mn-Mn distance (ab plane)/ Å** | 3.899 (2) | 3.896 (2) | 3.903 (2) |
| **Interlayer Mn-Mn distance/ Å** | 6.867 (4) | 6.931 (2) | 7.091 (3) |
| **Volume / Å$^3$** | 104.42 (9) | 105.25 (1) | 109.58 (5) |
| **Mn-Pt/Pd distance in Mn@(Pd/Pt)$_{12}$/Å** | | | |
| ***ab* plane** | 2.757 (3) | 2.755 (1) | 2.779 (2) |
| **Out of *ab* plane** | 2.810 (2) | 2.807 (1) | 2.790 (1) |
| **P-Pd/Pt/ Å** | 2.405 (2) | 2.425 (1) | 2.512 (1) |
| **Magnetic ordering** | FM | AFM | FM |
| **Transition temperature/ K** | 299 (2) | 188 (2) | 300 (2) |

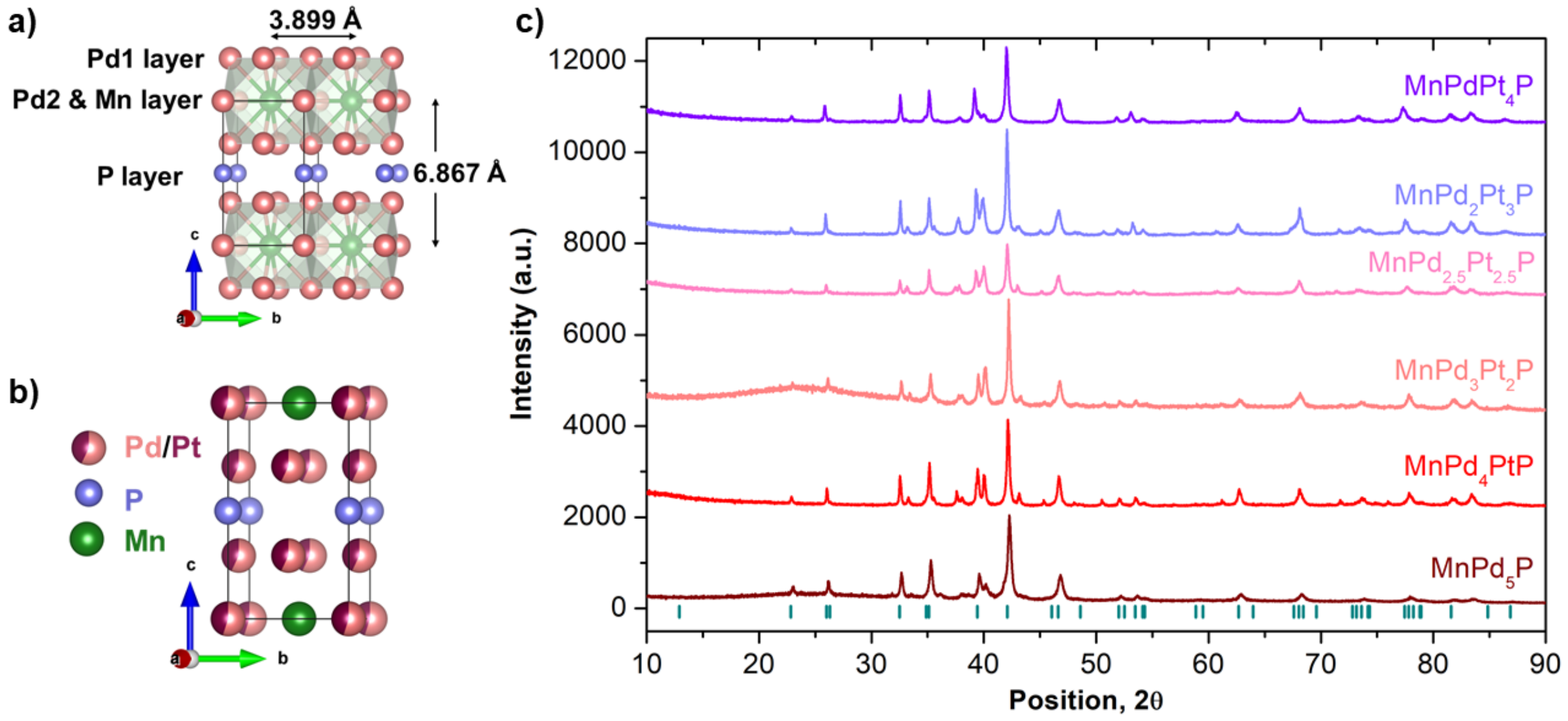

**Figure 2.** (***a***) The crystal structure of $MnPd_5P$ showing Mn@$Pd_{12}$ face sharing polyhedral and P layers (***b***) Unit cell of $MnPt_{5-x}Pd_xP$ indicating the mixture of Pd and Pt. (***c***) Powder diffraction patterns of $MnPt_{5-x}Pd_xP$ ($x$ = 1, 2, 2.5, 3, 4, and 5). The green vertical bars mark the expected Bragg positions for $MnPd_5P$.

Pd substitutions on $MnPt_5P$ yielded five new compounds with nominal compositions of $MnPt_{5-x}Pd_xP$ ($x$ = 1, 2, 2.5, 3, 4). These compounds were structurally characterized to maintain the parent lattice structure with different distribution of Pt/Pd mixture on the two atomic sites 1*a* and 1*b* as indicated in **Figure 2*b***. Details on the Pt/Pd distributions on 1*a* and 1*b* sites for each phase are given in the **Table S2.** The single crystal refinement results on Pd doped samples indicated some residual intensities on the 1*a* and 1*b* sites even after mixed occupancy of Pt/Pd is accounted. This observation suggests a possibility of Pd-Pt solid solution like nature on these materials rather than a completely ordered structure. This also provides an explaination for the refined high diffraction peak and hole values and goodness of fit for some of the $MnPt_{5-x}Pd_xP$ single crystal data which could originate from mismatch between the model and the actual structure. Moreover, synthesizing high quality single crystals for these compounds were a challenge and for $MnPdPt_4P$ irrespective of multiple synthesis attempts, we were unable to grow high quality single crystals to collect data to refine the structure, as a result only the lattice parameters are reported in **Table S1**. However, the SCXRD results, and chemical compositions can be further supported by the PXRD and SEM-EDX analysis. A comparison of PXRD for $MnPt_{5-x}Pd_xP$ ($x$ = 1, 2, 2.5, 3, 4, and 5) phases with the calculated Bragg peaks for the parent $MnPd_5P$ phase are presented in **Figure 2*c***. The LeBail fitting of PXRD for $MnPd_5P$ is given in the **Figure S1**. The refined lattice parameters for $MnPd_5P$ are in well agreement with the single crystal refinement data ($\chi^2$ = 1.99). PXRD analysis shows the high purity of the bulk $MnPd_5P$ phase. The possible impurity phases of Pd and $MnPd_3$ were tested in the refinement and confirmed the synthesis of pure phase $MnPd_5P$. The percentage of impurities in $MnPt_{5-x}Pd_xP$ samples were evaluated from the Rietveld refinement in the HighScore Plus software. Results indicate that loaded $MnPd_4PtP$, $MnPd_3Pt_2P$, $MnPd_{2.5}Pt_{2.5}P$, $MnPd_2Pt_3P$ and $MnPdPt_4P$ reactions yielded 86.8%, 82.7%, 51.2%, 75.8% and 93.1% of the targeted phase as the major phase respectively. The Cubic PdPt impurity was observed in all $MnPt_{5-x}Pd_xP$ ranging from 2-20% which further supports our speculation from SCXRD results about the Pd-Pt solid solution nature in these materials. However, the magnetic impurities MnPdP and MnPt were found to consist less than 5% in all doped samples except for $MnPd_{2.5}Pt_{2.5}P$ which was 32%. Detailed Rietveld refinement analysis with impurities is given in **Table S2**.

The chemical compositions and the stoichiometries were further evaluated using SEM-EDS and details are reported in **Table S3**. The chemical composition determined by SEM-EDS gives $Mn_{1.00}Pd_{4.73}P_{0.79}$, $Mn_{1.00}Pd_{3.99}Pt_{0.96}P_{0.88}$, $Mn_{1.00}Pd_{2.88}Pt_{2.01}P_{0.84}$, $Mn_{1.00}Pd_{2.48}Pt_{2.46}P_{0.83}$, $Mn_{1.00}Pd_{1.82}Pt_{2.82}P_{0.78}$, and $Mn_{1.00}Pd_{0.92}Pt_{3.96}P_{0.82}$ for the loaded Pt mol% of 0%, 14.2%, 28.5%, 35.7%, 42.8% and 57.1% respectively. Furthermore, these results are in good agreement with the single crystal results given in **Table 1 and Table S1**.

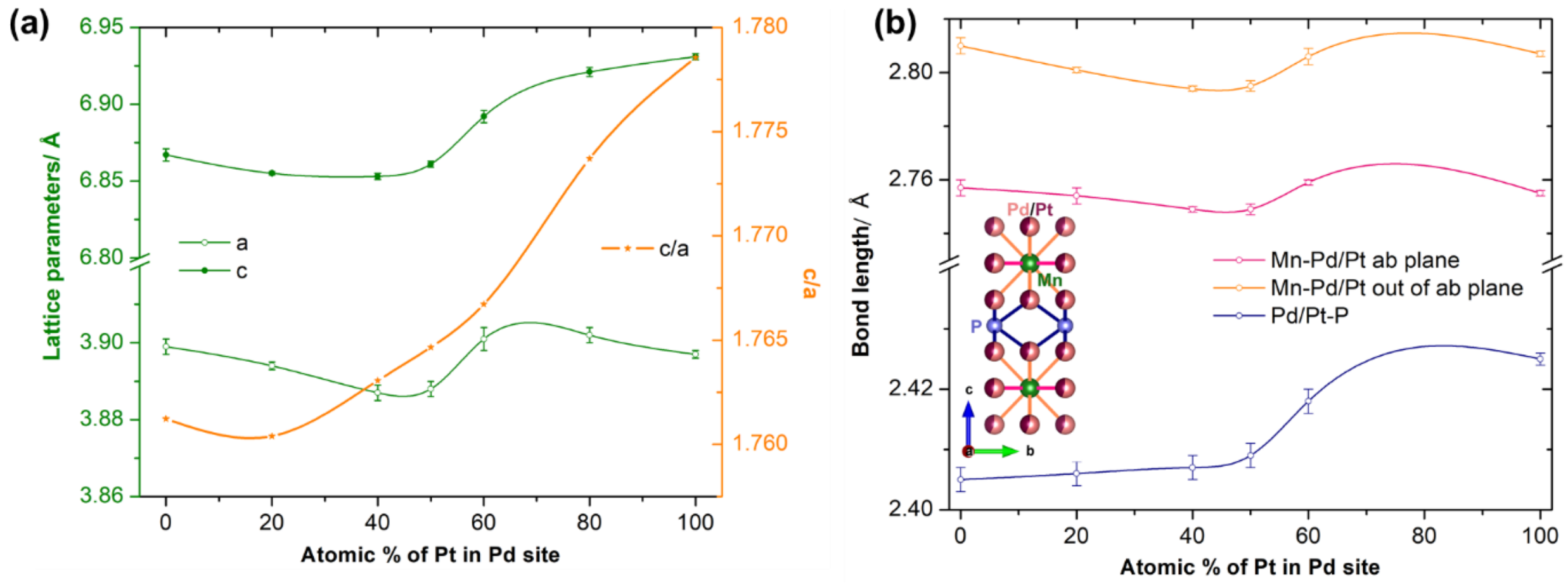

**Figure 3.** The comparison of structural features of $MnPt_{5-x}Pd_xP$ ($x$ = 1, 2, 3, 4 and 5) with $MnPt_5P$ showing **(*a*)** change of lattice parameters and *c*/*a* ratios **(b)** change of atomic bond distances within $Mn@(Pt/Pd)_{12}$ with changing Pd content.

A comparison of the structural features of the new anti-115-Pd variants with the related $MnPt_5P$ structure are given in **Figure 3.** As shown in **Figure 3*a*,** the lattice parameter *a* is almost constant (3.899 ± 0.004) in all compounds consequently leading to similar Mn-Mn distance in the *ab* plane while the *c* parameter of the new compounds $MnPt_{5-x}Pd_xP$ has shortened ~1% compared to $MnPt_5P$. Moving from 5*d* (Pt) to 4*d* (Pd) has not significantly changed the *ab*-plane of the structure due to the close atomic size of Pt to Pd caused by the lanthanide contraction and shielding of Pt. Overall, the $Mn@(Pd/Pt)_{12}$ shows similar features in all compounds as indicated in **Figure 3*b*.** Interestingly, the *c* axis is sensitive to the changes in chemical composition leading to different magnetic properties in each compound, ranging from canted antiferromagnetic $MnPt_5P$ to ferromagnetic $MnPd_5P$ in this family.

**Vanishing Spin-Canting by Replacing Pt with Pd:** The spin-canted antiferromagnetic ordering was detected in $MnPt_5P$ with the $T_N$ ~ 188 K by our previous work.[10] However, at 350 K, the ferromagnetic component along the *c*-axis still exists in $MnPt_5P$. By replacing Pt with Pd up to 40%, the ferromagnetic transition appears at ~ 354 K (20% Pd) and ~ 366 K (40% Pd), while the spin-canting temperature increases to ~ 246 K (20% Pd) and ~ 299 K (40% Pd) respectively. Once the Pd replaces 50% of Pt, the spin-canting disappears, and only the ferromagnetic transition was observed around 300 K. The polycrystalline samples of $MnPt_{5-x}Pd_xP$ were used to measure the magnetic properties with an external magnetic field of 1000 Oe from 1.8 K to 350 K/ 400 K or 600 K as required. The zero-field cooling (ZFC) and field cooling (FC) graphs of the magnetic susceptibility (χ) vs temperature collected below 350 K are shown in the **Figure 4** q χ vs T for all compounds are given in **Figure S2**. Upon cooling each sample, a dramatic increase in magnetic susceptibility was observed around room temperature or above suggesting the ferromagnetic ordering of all $MnPt_{5-x}Pd_xP$ compounds. The transition temperature ($T_c$) for each composition is plotted against the atomic % of Pt in each material and given in the **Figure 5** which are ranging from ~299 to ~366 K. The $T_c$ of each compound was determined using the derivative of χ vs T. As indicated in the **Figure 5**, the $T_c$ of these materials increase with increasing Pt content, reach the maximum around 30% before it starts decreasing around 40%. Starting from 300 K in $MnPd_5P$, $T_c$ of $MnPd_4PtP$ increased to about 316 K before it starts lowering from $MnPd_3Pt_2P$ and onwards. Furthermore, when the Pt content is above 50 %, two transitions were observed for $MnPd_2Pt_3P$ and $MnPdPt_4P$ compounds. $T_{c1}$ and $T_{c2}$ for $MnPd_2Pt_3P$ are found to be 366 K and 299 K while for $MnPdPt_4P$, it was 354 K and 246 K respectively. Furthermore, in $MnPt_5P$, AFM transition $T_N$ ~ 188 K is reported while results evident that FM interactions along *c* axis are still there above 350 K as χ has not reached zero.[10] Therefore, for the FM transition in $MnPt_5P$, a hypothetical value above 350 K has used to depict how the two transitions from canted AFM merge in to the purely FM transition in $MnPd_5P$ with increasing Pd content in **Figure 5**. This indicates that by lowering the Pt content, the gap between $T_{c1}$ and $T_{c2}$/ $T_N$ has lowered upto 40 % Pt, merging to one transition around 50 % Pt disappearing the AFM interactions in the $MnPt_{5-x}Pd_xP$ system. Moreover, when the materials with single $T_c$, $MnPt_{5-x}Pd_xP$ ($x$ = 2.5, 3, 4, 5) are considered, it is revealed that the system has the highest $T_c$ when *c/a* ratio is minimum while the highest *c/a* ratio yielded an antiferromagnet $MnPt_5P$. Similar magnetic crossovers from AFM to FM has been observed in Mn sublattice in $YbMn_6Sn_6$ system with $HfFe_6Ge_6$ type structure where tuning the valance electron concentrations through chemical substitution of Yb site drives this magnetic behavior.[23–25] In $SrCo_2(Ge_{1-x}P_x)_2$ system where both $SrCo_2P_2$ and $SrCo_2Ge_2$ are paramagnetic, weak itinerant ferromagnetism has developed by the use of chemical bond breaking as a tuning parameter.[26–28] Additionally, introduction of Eu into AFM $PrCo_2P_2$ has pushed the material into an itinerant FM $Pr_{0.8}Eu_{0.2}Co_2P_2$ due to a strong chemical compression.[29,30] It is clear that magnetic crossovers in the systems can be manipulated through various parameters while the observed AFM to FM transformation in $MnPt_{5-x}Pd_xP$ system is driven by tuning the unit cell dimensions and spin-orbit coupling effects.

A significant difference in the ZFC and FC measurements at 1000 Oe was not observed in $MnPd_5P$ while all Pt doped compounds indicated a split of ZFC, and FC measurements as shown in **Figure 4*a-f***. Previous studies have shown that in some ferromagnetic systems such as perovskites $SrRuO_3$, $La_{0.7}Ca_{0.3}MnO_3$ and $La_{0.5}Sr_{0.5}CoO_3$, the irreversibility indicated by the difference in ZFC and FC magnetic susceptibilities result from magnetic anisotropy.[31] Accordingly, when using this as a measure of magnetic anisotropy, it is clear that $MnPd_5P$ shows minimum magnetic anisotropy while all Pt doped variants possesses magnetic anisotropy to varying extents in the system as a result of higher spin-orbit coupling contribution from Pt.

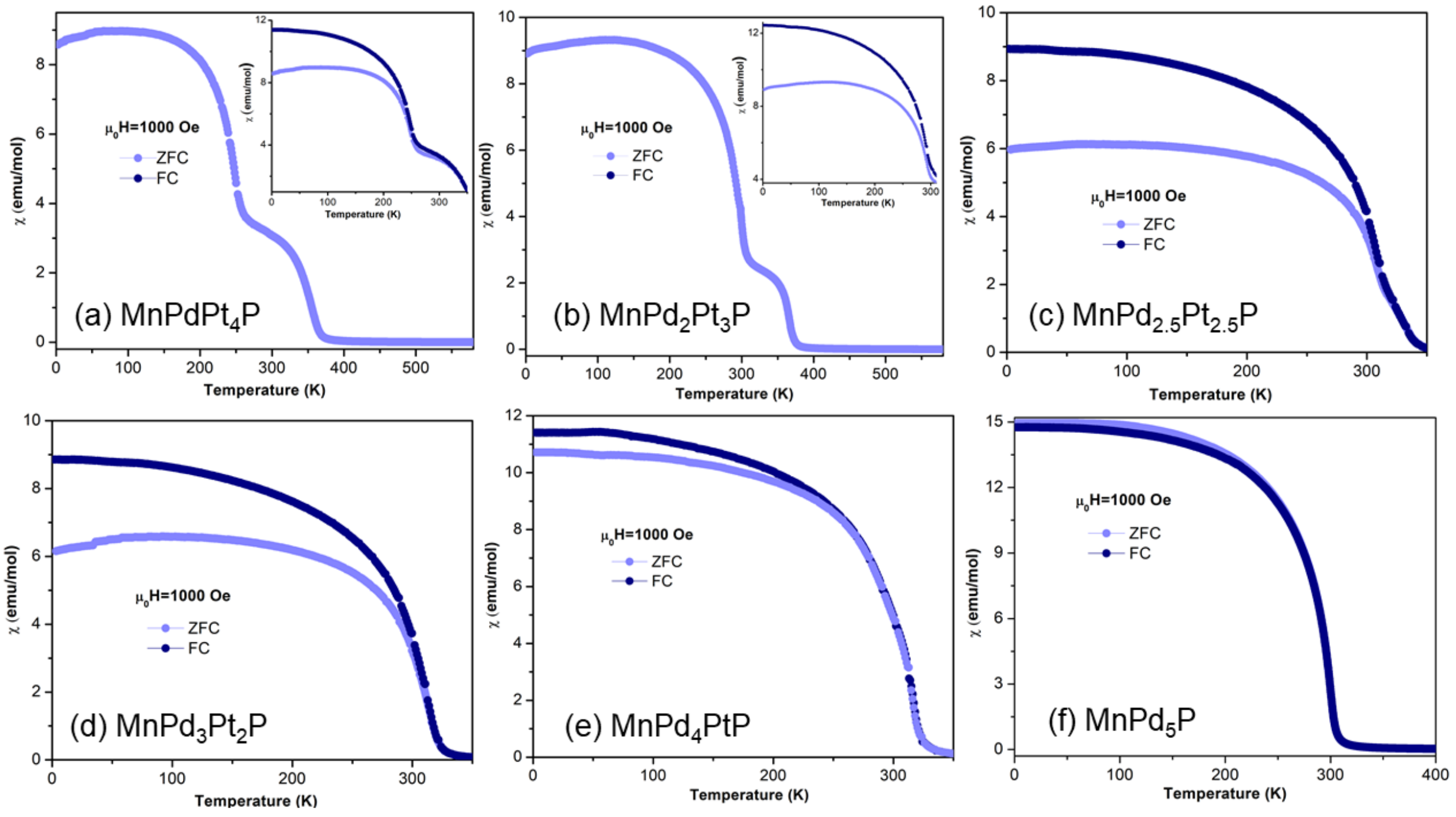

**Figure 4.** Magnetic susceptibility versus temperature measured by zero field cooled (ZFC) and field cooled (FC) methods at 1000 Oe for ***(a)*** $MnPdPt_4P$ ***(b)*** $MnPd_2Pt_3P$ ***(c)*** $MnPd_{2.5}Pt_{2.5}P$ ***(d)*** $MnPd_3Pt_2P$ ***(e)*** $MnPd_4PtP$ and ***(f)*** $MnPd_5P$. For ***(a)*** and ***(b)*** the main panel shows the combined ZFC $\chi$ measurements at low temperatures and high temperatures while the insets represent ZFC and FC graphs measured below room temperature.

The high temperature region of the magnetic susceptibility data was analyzed by and the Curie-Weiss law, $\chi = \frac{C}{T-\theta_{CW}}$ and the fitting for each material is given in **Figure S3.** The symbols, $\chi$, $C$, and $\theta_{CW}$ represent magnetic susceptibility, temperature independent contribution to the susceptibility, Curie constant and the Curie-Weiss temperature, respectively. The effective magnetic moments ($\mu_{eff}$ per Mn atom) ($\mu_{eff} = \sqrt{8c}\,\mu_B$) calculated from Curie-Weiss fitting is given in in the **Table 4**. $MnPd_5P$ indicate a higher $\mu_{eff}$ ~ $5\mu_B$ than ~$3.5\mu_B$ observed in ferromagnetically ordered $MnPt_5As$ from the same family.[11] The fitted Curie-Weiss temperatures are in a close agreement with the observed ferromagnetic transitions for each composition.

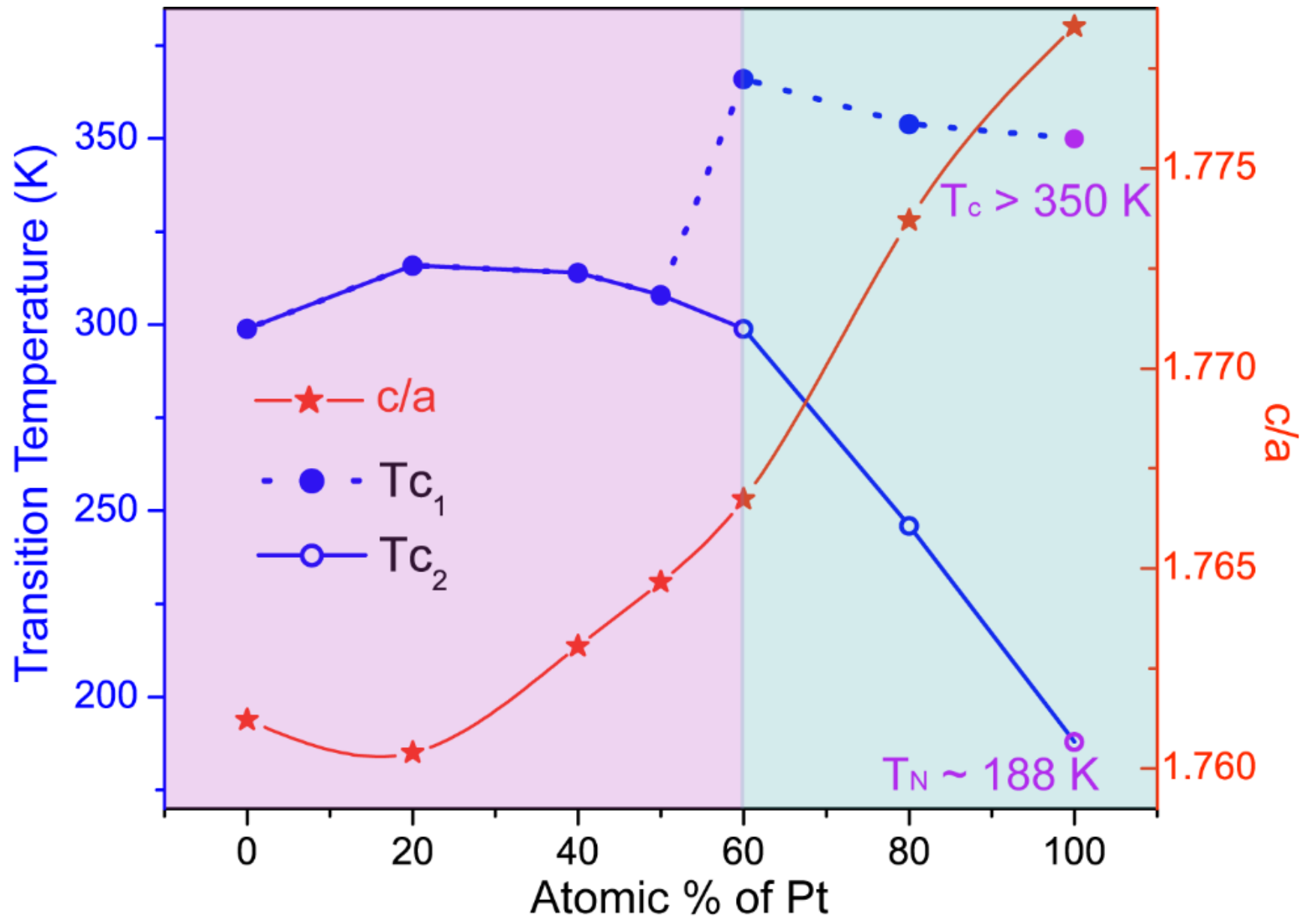

**Figure 5.** The trend of transition temperatures of $MnPd_5P$, $MnPt_{5-x}Pd_xP$ and $MnPt_5P$ relative to the change of *c/a* ratio.

The field dependent magnetization measurement of the $MnPt_{5-x}Pd_xP$ ($x$ = 1, 2, 2.5, 3, 4, 5) is shown in **Figure 6**. The hysteresis loops of magnetization were measured at different temperatures ranging from 2 K to 350 K with the external fields up to 90 kOe. The isothermal magnetization curves at 2 K tend to saturate around ~$4\mu_B$ /Mn by the applied external field of ~1 T for all compounds. With the increasing temperature saturated magnetic moments gradually decreases upto room temperature (300 K) and disappears above 325 K for $MnPt_{5-x}Pd_xP$ ($x$ = 1, 2, 2.5, 3, 4, 5). However, for compounds $MnPd_2Pt_3P$ and $MnPdPt_4P$ magnetic moments tend to saturate upto 350 K. These results further support that all $MnPt_{5-x}Pd_xP$ ($x$ = 1, 2, 2.5, 3, 4, 5) compounds are ferromagnets above room temperature. The coercivity of each compound at 2 K is presented in **Figure 7a**. The results indicate that all $MnPt_{5-x}Pd_xP$ ($x$ = 1, 2, 2.5, 3, 4, 5) compounds can be categorized into semi-hard ferromagnets as their coercivities are in the range 13 Oe – 376 Oe.[5] This is unlike the soft ferromagnetism observed in $MnPt_5As$.[11] The change of coercivities with respect to Pt content in the structure is represented in the ***Inset*** of **Figure 7** suggesting a non-linear dependence. The critical fields and maximum magnetization derived from isothermal magnetization curves at 2 K are shown in **Figure 7b** which indicates a non-linear dependence in $MnPt_{5-x}Pd_xP$ system unlike for the Mn sublattice in $YbMn_6Sn_6$ system.[23,25]

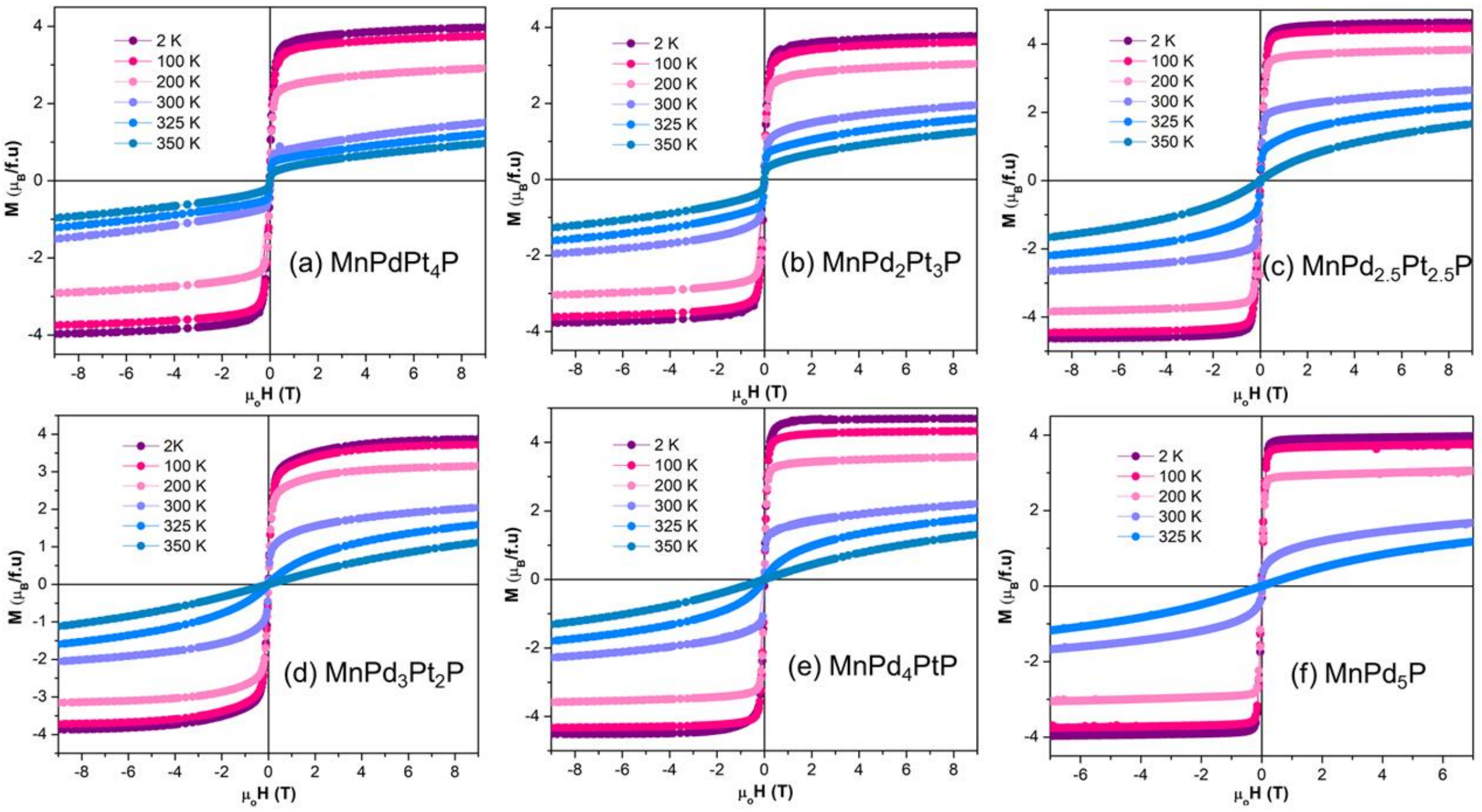

**Figure 6.** Magnetization vs applied field at temperatures of 2, 100, 200, 300, 325 and 350 K for ***(a)*** $MnPt_4PdP$ ***(b)*** $MnPt_3Pt_2P$ ***(c)*** $MnPt_{2.5}Pd_{2.5}P$ ***(d)*** $MnPt_2Pd_3P$ ***(e)*** $MnPtPd_4P$ and ***(f)*** $MnPd_5P$

**Table 4.** Summary of magnetic parameters derived from Curie-Weiss fitting

| | $MnPd_5P$ | $MnPd_4PtP$ | $MnPd_3Pt_2P$ | $MnPd_{2.5}Pt_{2.5}P$ | $MnPd_2Pt_3P$ | $MnPdPt_4P$ | $MnPt_5P$ |
|---|---|---|---|---|---|---|---|
| $T_N$ (K) | / | / | / | / | 299 | 246 | 188 |
| $T_c$ (K) | 300 | 316 | 314 | 308 | 366 | 354 | >350 |
| $\theta_{CW}$ (K) | 304.4(1) | 317.8(3) | 318.0(1) | 334.1(2) | 369.9(1) | 366.2(3) | |
| C | 3.13(1) | 4.14(6) | 2.45(1) | 2.09(4) | 1.09(1) | 1.16(1) | |
| $\mu_{eff}$ ($\mu_B$) | 5.00 | 5.75 | 4.43 | 4.09 | 2.95 | 3.04 | |
| Coercivity (Oe) | ~93 | ~68 | ~91 | ~77 | ~52 | ~68 | |

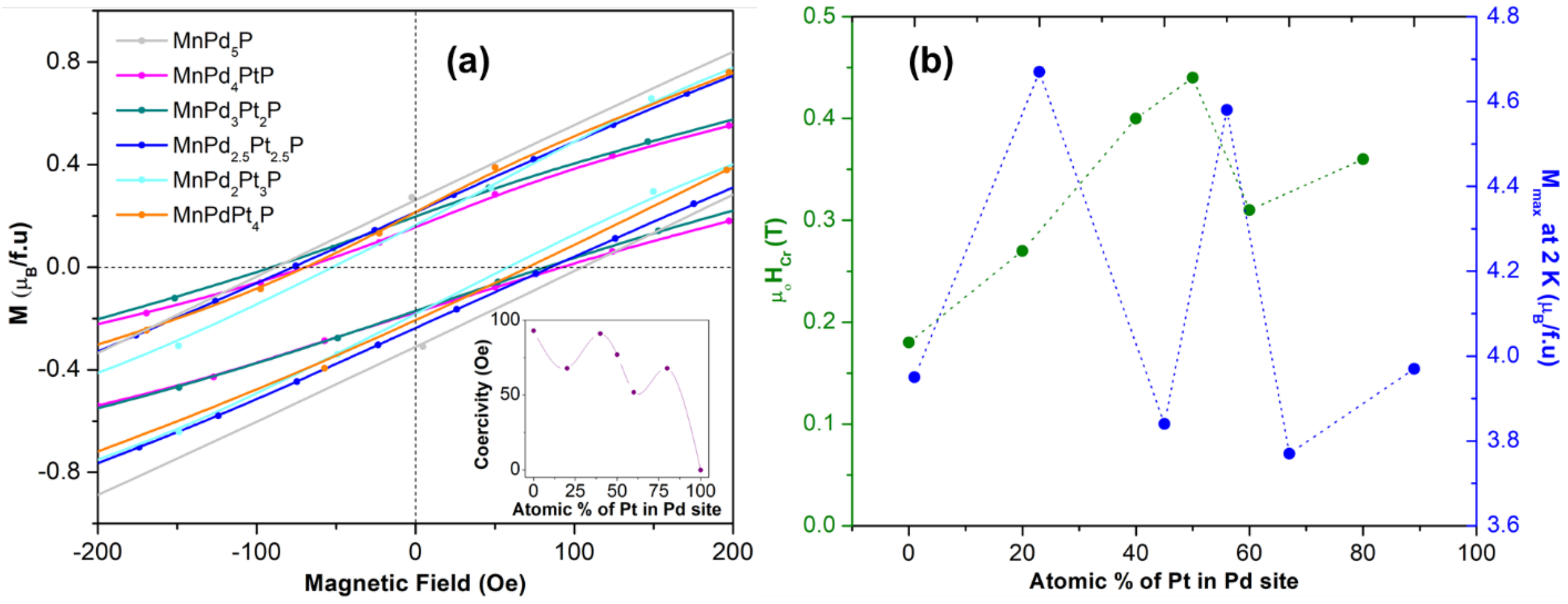

**Figure 7. (a)** Hysteresis loops measured under low field at 2 K for each compound $MnPt_{5-x}Pd_xP$ ($x$ = 1, 2, 2.5, 3, 4, 5). ***Inset*** showing the change of coercivity of with the Pt content in the anti-115 compounds. **(b)** Critical field ($\mu_o H_{Cr}$) on the left axis and maximum magnetization on the right axis extracted from isothermal magnetization at 2 K for $MnPt_{5-x}Pd_xP$ ($x$ = 1, 2, 2.5, 3, 4, 5). The lines are only eye guides.

## Conclusion

We have successfully synthesized the family compounds $MnPt_{5-x}Pd_xP$ ($x$ = 1, 2, 2.5, 3, 4, 5) using high temperature solid state synthesis method, characterized the structural and magnetic properties of the materials. $MnPd_5P$ in the anti-$CeCoIn_5$-system was theoretically predicted and experimentally confirmed to show stable ferromagnetic properties. $MnPd_5P$ is a low dimensional itinerant ferromagnet with a Curie-Weiss temperature of ~305 K and semi-hard magnetic nature. This study provides a new platform in addition to the widely studied $CeCoIn_5$-structure type to understand the tunability of structure and magnetism and predict new rare-earth free magnets.

## Acknowledgements

The work at Rutgers is supported by Beckman Young Investigator award and NSF-DMR-2053287. C. -J.K. and G.K. were supported by the U.S. Department of Energy, Office of Science (Basic Energy Science) as a part of the Computational Materials Science Program through the Center for Computational Design of Functional Strongly Correlated Materials and Theoretical Spectroscopy under DOE grant no. DE-FOA-0001276. C.-J.K. also acknowledges support by NRF grant No. 2022R1C1C1008200. A portion of this work was performed at the National High Magnetic Field Laboratory, which is supported by National Science Foundation Cooperative Agreement No. DMR-1644779 and the State of Florida.

## Conflicts of interest

The authors declare that there is no conflict of interest.

# Supplementary Information

## Theory-guided Investigation on Magnetic Evolution of $MnPt_{5-x}Pd_xP$ and Discovery of anti-$CeCoIn_5$-type Ferromagnetic $MnPd_5P$

*Ranuri S. Dissanayaka Mudiyanselage,[1] Chang-Jong Kang,[2,3] Kaya Wei,[4] Zhixue Shu[5], Tai Kong[5], Ryan Baumbach,[4] Gabriel Kotliar,[2] Weiwei Xie[1]**

*[1]Department of Chemistry and Chemical Biology, Rutgers University, Piscataway, NJ, 08854, USA*
*[2]Department of Physics and Astronomy, Rutgers University, Piscataway, NJ, 08854, USA*
*[3]Department of Physics, Chungnam National University, Daejeon, 34134, South Korea*
*[4]National High Magnetic Field Laboratory, Tallahassee, FL, 32310, USA*
*[5]Department of Physics, University of Arizona, Tucson, AZ 85721, USA*

# Table of Contents

**Table S1.** Single crystal structure refinement for $MnPt_{5-x}PdP$ at 300 (2) K. (Standard deviation is indicated by the values in parentheses)

| **Loaded composition** | **$MnPd_5P$** | **$MnPd_4PtP$** | **$MnPd_3Pt_2P$** | **$MnPd_{2.5}Pt_{2.5}P$** | **$MnPd_2Pt_3P$** | **$MnPdPt_4P$** |
|---|---|---|---|---|---|---|
| Refined Formula | $MnPd_5P$ | $MnPd_{4.14(2)}Pt_{0.86(2)}P$ | $MnPd_{2.73}Pt_{2.27}P$ | $MnPd_{2.4}Pt_{2.6}P$ | $MnPd_{2.10}Pt_{2.90}P$ | |
| F.W. (g/mol) | 617.91 | 694.18 | 819.24 | 848.50 | 875.11 | |
| Space group; Z | *P* 4/*mmm*; 1 | *P* 4/*mmm*; 1 | *P* 4/*mmm*; 1 | *P* 4/*mmm*; 1 | *P* 4/*mmm*; 1 | |
| *a*(Å) | 3.899 (2) | 3.894 (1) | 3.887 (2) | 3.888 (2) | 3.901 (3) | 3.902(2) |
| *c*(Å) | 6.867 (4) | 6.855 (1) | 6.853 (2) | 6.861 (2) | 6.892 (4) | 6.921(3) |
| V (Å$^3$) | 104.42 (2) | 103.93 (2) | 103.54 (4) | 103.73 (5) | 104.93 (11) | 105.41 (7) |
| θ range (°) | 2.966-34.770 | 5.236-34.828 | 5.246-34.892 | 2.969-3.874 | 5.919-34.646 | |
| No. reflections; $R_{int}$ | 578; 0.0609 | 1397; 0.0293 | 999; 0.0444 | 904; 0.0549 | 214; 0.0233 | |
| No. independent reflections | 170 | 173 | 166 | 167 | 94 | |
| No. parameters | 12 | 14 | 14 | 14 | 14 | |
| $R_1$: $\omega R_2$ ($I$>2□($I$)) | 0.0509; 0.1204 | 0.0253; 0.0603 | 0.0429; 0.1026 | 0.0417; 0.1080 | 0.0373; 0.0947 | |
| Goodness of fit | 1.282 | 1.346 | 1.331 | 1.429 | 1.154 | |
| Diffraction peak and hole (e$^-$/ Å$^3$) | 2.656; -1.863 | 2.632; -1.926 | 7.771; -3.765 | 7.323; 5.219 | 2.579; -3.540 | |
| Temperature | 300 (2) | 299 (2) | 301 (2) | 300 (2) | 301 (2) | |

**Table S2.** Atomic coordinates, occupancies and equivalent isotropic displacement parameters of $MnPt_{5-x}PdP$ at 300 (2) K. ($U_{eq}$ is defined as one-third of the trace of the orthogonalized $U_{ij}$ tensor (Å$^2$)).

| **Atom** | **Wyckoff.** | **Occ.** | ***x*** | ***y*** | ***z*** | $U_{eq}$ |
|---|---|---|---|---|---|---|
| | | | MnPd5P | | | |
| Pd1 | 4*i* | 1 | 0 | ½ | 0.2948 (1) | 0.015(1) |
| Pd2 | 1*a* | 1 | 0 | 0 | 0 | 0.012(2) |
| Mn3 | 1*c* | 1 | ½ | ½ | 0 | 0.022(1) |
| P4 | 1*b* | 1 | 0 | 0 | ½ | 0.016(2) |
| | | | $MnPd_4PtP$ | | | |
| Pd1 | 4*i* | 0.84(2) | 0 | ½ | 0.29385(9) | 0.0058(2) |
| Pt2 | 4*i* | 0.22(2) | 0 | ½ | 0.2948 (1) | 0.0058(2) |
| Pd3 | 1*a* | 0.78(2) | 0 | 0 | 0 | 0.0046(3) |
| Pt4 | 1*a* | 0.22(2) | 0 | 0 | 0 | 0.0046(3) |
| Mn3 | 1*c* | 1 | ½ | ½ | 0 | 0.0128(9) |
| P4 | 1*b* | 1 | 0 | 0 | ½ | 0.0076(11) |
| | | | $MnPd_3Pt_2P$ | | | |
| Pd1 | 4*i* | 0.57(6) | 0 | ½ | 0.29296(17) | 0.0056(4) |
| Pt2 | 4*i* | 0.43 (6) | 0 | ½ | 0.29296(17) | 0.0056(4) |
| Pd3 | 1*a* | 0.55(6) | 0 | 0 | 0 | 0.0037(6) |
| Pt4 | 1*a* | 0.45(6) | 0 | 0 | 0 | 0.0037(6) |
| Mn3 | 1*c* | 1 | ½ | ½ | 0 | 0.020(2) |
| P4 | 1*b* | 1 | 0 | 0 | ½ | 0.006(2) |
| | | | $MnPd_{2.5}Pt_{2.5}P$ | | | |
| Pd1 | 4*i* | 0.50(8) | 0 | ½ | 0.29274(18) | 0.0078(5) |
| Pt2 | 4*i* | 0.50 (8) | 0 | ½ | 0.29274(18) | 0.0078(5) |
| Pd3 | 1*a* | 0.42(6) | 0 | 0 | 0 | 0.0049(7) |
| Pt4 | 1*a* | 0.58(6) | 0 | 0 | 0 | 0.0049(7) |
| Mn3 | 1*c* | 1 | ½ | ½ | 0 | 0.015(3) |
| P4 | 1*b* | 1 | 0 | 0 | ½ | 0.011(3) |
| | | | $MnPd_2Pt_3P$ | | | |
| Pd1 | 4*i* | 0.43(6) | 0 | ½ | 0.29254(11) | 0.0084(6) |
| Pt2 | 4*i* | 0.57 (6) | 0 | ½ | 0.29254(11) | 0.0084(6) |
| Pd3 | 1*a* | 0.38(6) | 0 | 0 | 0 | 0.0068(7) |
| Pt4 | 1*a* | 0.62(6) | 0 | 0 | 0 | 0.0068(7) |
| Mn3 | 1*c* | 1 | ½ | ½ | 0 | 0.012(3) |
| P4 | 1*b* | 1 | 0 | 0 | ½ | 0.014(4) |

**Figure S1.** Powder diffraction pattern of $MnPd_5P$. The red, black and blue solid line represent experimental data ($I_{obs}$), the calculated LeBail refinement ($I_{cal}$) and the difference between calculated and observed intensities ($I_{obs} - I_{cal}$), respectively. The green vertical bars mark the expected Bragg positions for $MnPd_5P$.

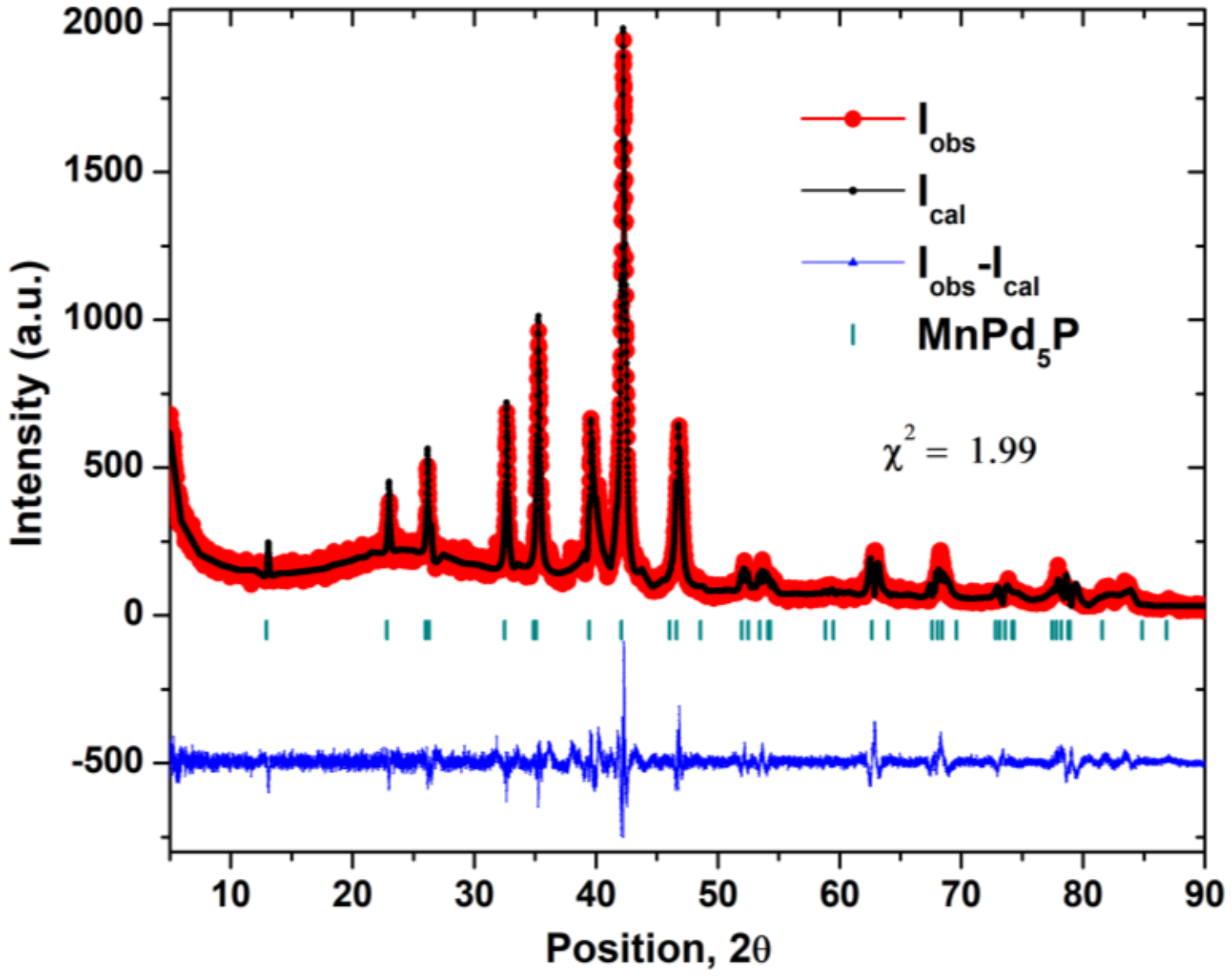

**Table S3.** The percentages of impurities determined from the Rietveld refinement in the HighScore Plus software for $MnPt_{5-x}Pd_xP$.

| | $MnPt_{5-x}Pd_xP$ % | PdPt % | MnPdP % | Pt % | MnPt % |
|---|---|---|---|---|---|
| $MnPd_4PtP$ | 86.8 | 9.8 | 3.1 | 0.3 | 0 |
| $MnPd_3Pt_2P$ | 82.7 | 14.4 | 3.2 | 1.7 | 0.6 |
| $MnPd_{2.5}Pt_{2.5}P$ | 51.2 | 15.4 | 32 | 1.4 | 0 |
| $MnPd_2Pt_3P$ | 75.8 | 19.1 | 4.1 | 1.0 | 0 |
| $MnPdPt_4P$ | 93.1 | 2.0 | 1.5 | 3.3 | 0 |

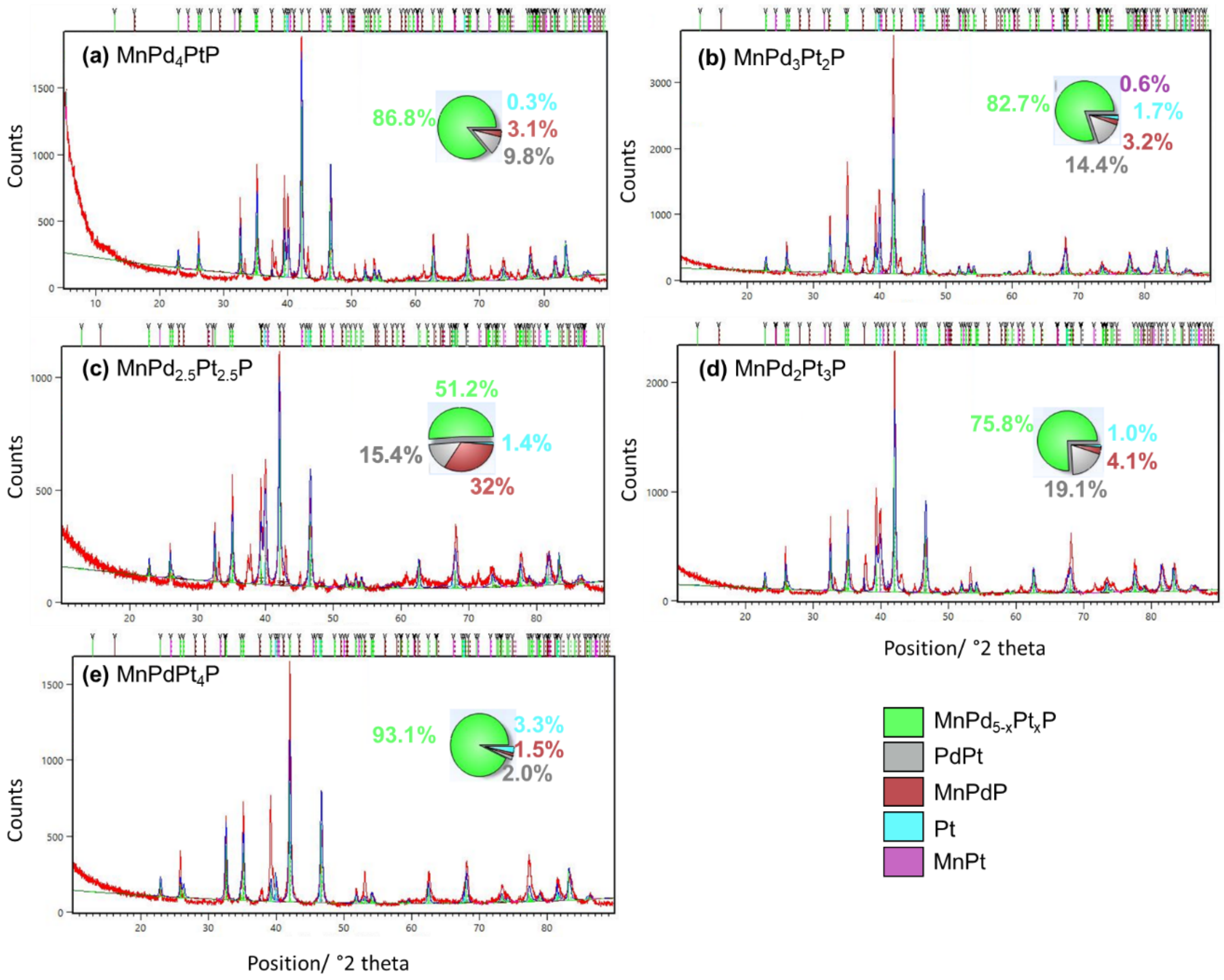

**Table S4.** Chemical compositions obtained from SEM-EDX for the loaded $MnPt_{5-x}Pd_xP$ samples and the corresponding crystal images

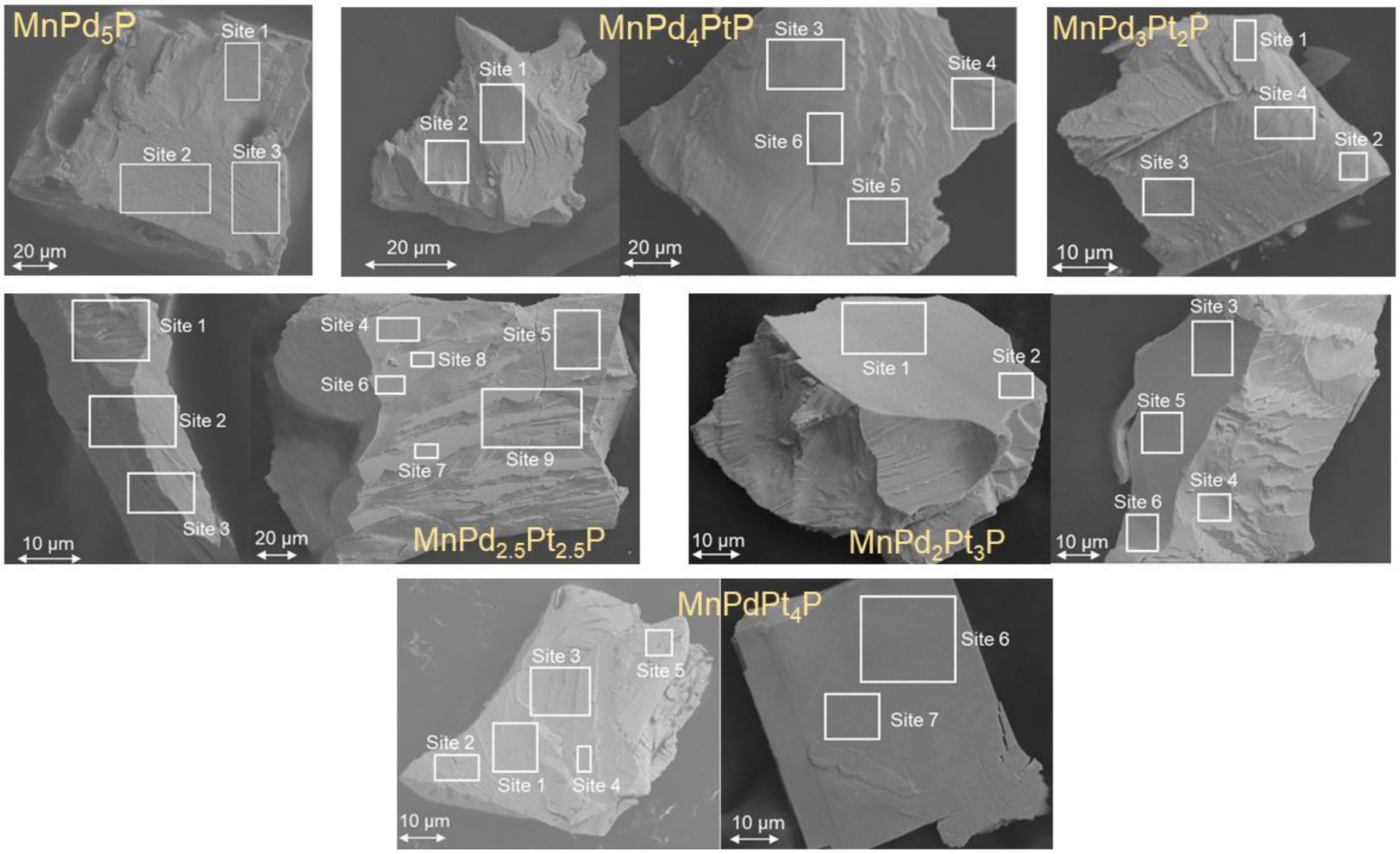

| **$MnPd_5P$** | | | | | | |
|---|---|---|---|---|---|---|
| | | *Mn* | *Pd* | *P* | | *Composition* |
| *Crystal-1* | *Site1* | 14.95 | 72.34 | 12.71 | | $Mn_{1.00}Pd_{4.84}P_{0.85}$ |
| | | | | | | $Mn_{1.05}Pd5.06P_{0.89}$ |
| | | | | | | $Mn_{1.06}Pd5.09P0.85$ |
| | | | | | | $Mn_{1.12}Pd5.10P_{0.81}$ |
| | | | | | | $Mn_{1.07}Pd_{5.08}P_{0.85}$ |
| | *Site2* | 15.10 | 72.78 | 12.12 | | $Mn_{1.00}Pd_{4.82}P_{0.80}$ |
| | *Site3* | 15.95 | 72.38 | 11.67 | | $Mn_{1.00}Pd_{4.54}P_{0.73}$ |
| | | | | | ***Average*** | $\mathbf{Mn_{1.00}Pd_{4.73}P_{0.79}}$ |
| **$MnPd_4PtP$** | | | | | | |
| | | *Mn* | *Pd* | *Pt* | *P* | *Composition* |
| *Crystal-1* | *Site1* | 14.60 | 59.07 | 13.03 | 13.29 | $Mn_{1.00}Pd_{4.05}Pt_{0.89}P_{0.91}$ |
| | *Site2* | 17.5 | 57.00 | 12.91 | 12.59 | $Mn_{1.00}Pd_{3.26}Pt_{0.74}P_{0.72}$ |
| *Crystal-2* | *Site3* | 13.99 | 58.20 | 15.12 | 12.68 | $Mn_{1.00}Pd_{4.16}Pt_{1.08}P_{0.91}$ |
| | *Site4* | 13.95 | 60.65 | 12.99 | 12.40 | $Mn_{1.00}Pd_{4.35}Pt_{0.93}P_{0.89}$ |
| | *Site5* | 14.33 | 58.28 | 14.96 | 12.43 | $Mn_{1.00}Pd_{4.07}Pt_{1.04}P_{0.87}$ |
| | *Site6* | 14.01 | 57.14 | 15.14 | 13.47 | $Mn_{1.00}Pd_{4.08}Pt_{1.08}P_{0.96}$ |
| | | | | | ***Average*** | $\mathbf{Mn_{1.00}Pd_{3.99}Pt_{0.96}P_{0.88}}$ |
| **$MnPd_3Pt_2P$** | | | | | | |
| | | *Mn* | *Pd* | *Pt* | *P* | *Composition* |
| *Crystal-1* | *Site1* | 15.09 | 42.82 | 29.91 | 12.18 | $Mn_{1.00}Pd_{2.84}Pt_{1.98}P_{0.81}$ |
| | *Site2* | 15.14 | 41.98 | 29.98 | 12.63 | $Mn_{1.00}Pd_{2.77}Pt_{1.98}P_{0.83}$ |
| | *Site3* | 14.01 | 43.74 | 30.04 | 12.21 | $Mn_{1.00}Pd_{3.12}Pt_{2.14}P_{0.87}$ |
| | *Site4* | 15.35 | 42.50 | 29.44 | 12.71 | $Mn_{1.00}Pd_{2.77}Pt_{1.92}P_{0.83}$ |
| | | | | | ***Average*** | $\mathbf{Mn_{1.00}Pd_{2.88}Pt_{2.01}P_{0.84}}$ |
| **$\mathbf{MnPd_{2.5}Pt_{2.5}P}$** | | | | | | |
| | | *Mn* | *Pd* | *Pt* | *P* | *Composition* |
| *Crystal-1* | *Site1* | 13.81 | 37.78 | 36.89 | 11.53 | $Mn_{1.00}Pd_{2.74}Pt_{2.67}P_{0.83}$ |
| | *Site2* | 14.08 | 37.44 | 35.78 | 12.69 | $Mn_{1.00}Pd_{2.66}Pt_{2.54}P_{0.90}$ |
| | *Site3* | 14.71 | 37.31 | 35.94 | 12.04 | $Mn_{1.00}Pd_{2.54}Pt_{2.44}P_{0.82}$ |
| *Crystal-2* | *Site4* | 15.50 | 35.99 | 36.54 | 11.98 | $Mn_{1.00}Pd_{2.32}Pt_{2.36}P_{0.77}$ |
| | *Site5* | 13.83 | 37.42 | 36.72 | 12.03 | $Mn_{1.00}Pd_{2.71}Pt_{2.66}P_{0.87}$ |
| | *Site6* | 16.11 | 35.40 | 35.96 | 12.53 | $Mn_{1.00}Pd_{2.20}Pt_{2.23}P_{0.78}$ |
| | *Site7* | 15.53 | 35.99 | 36.56 | 11.93 | $Mn_{1.00}Pd_{2.32}Pt_{2.35}P_{0.77}$ |
| | *Site8* | 15.26 | 35.97 | 36.37 | 12.40 | $Mn_{1.00}Pd_{2.36}Pt_{2.38}P_{0.81}$ |
| | *Site9* | 14.44 | 36.41 | 36.50 | 12.65 | $Mn_{1.00}Pd_{2.52}Pt_{2.53}P_{0.88}$ |
| | | | | | ***Average*** | $\mathbf{Mn_{1.00}Pd_{2.48}Pt_{2.46}P_{0.83}}$ |
| **$\mathbf{MnPd_{2}Pt_{3}P}$** | | | | | | |
| | | *Mn* | *Pd* | *Pt* | *P* | *Composition* |
| *Crystal-1* | *Site1* | 16.37 | 28.07 | 44.39 | 11.17 | $Mn_{1.00}Pd_{1.71}Pt_{2.71}P_{0.68}$ |
| | *Site2* | 21.27 | 25.30 | 42.46 | 10.97 | $Mn_{1.00}Pd_{1.19}Pt_{2.00}P_{0.52}$ |
| *Crystal-2* | *Site3* | 13.07 | 30.26 | 44.22 | 12.45 | $Mn_{1.00}Pd_{2.32}Pt_{3.38}P_{0.95}$ |
| | *Site4* | 17.42 | 25.70 | 44.62 | 12.26 | $Mn_{1.00}Pd_{1.48}Pt_{2.56}P_{0.70}$ |
| | *Site5* | 13.56 | 30.28 | 44.25 | 11.90 | $Mn_{1.00}Pd_{2.23}Pt_{3.26}P_{0.88}$ |
| | *Site6* | 14.47 | 28.73 | 43.53 | 13.27 | $Mn_{1.00}Pd_{1.99}Pt_{3.01}P_{0.92}$ |
| | | | | | ***Average*** | $\mathbf{Mn_{1.00}Pd_{1.82}Pt_{2.82}P_{0.78}}$ |
| **$\mathbf{MnPdPt_{4}P}$** | | | | | | |
| | | *Mn* | *Pd* | *Pt* | *P* | *Composition* |
| *Crystal-1* | *Site1* | 14.30 | 13.88 | 58.83 | 13.00 | $Mn_{1.00}Pd_{0.97}Pt_{4.11}P_{0.91}$ |
| | *Site2* | 15.39 | 13.70 | 58.72 | 12.18 | $Mn_{1.00}Pd_{0.89}Pt_{3.81}P_{0.79}$ |
| | *Site3* | 14.22 | 14.11 | 58.93 | 12.75 | $Mn_{1.00}Pd_{0.99}Pt_{4.15}P_{0.90}$ |
| | *Site4* | 14.56 | 14.96 | 59.22 | 11.26 | $Mn_{1.00}Pd_{0.97}Pt_{4.11}P_{0.91}$ |
| | *Site5* | 15.61 | 13.82 | 58.83 | 11.75 | $Mn_{1.00}Pd_{0.88}Pt_{3.77}P_{0.76}$ |
| *Crystal-3* | *Site6* | 16.32 | 12.93 | 59.75 | 11.0 | $Mn_{1.00}Pd_{0.79}Pt_{3.66}P_{0.67}$ |
| | *Site7* | 14.59 | 13.77 | 59.95 | 11.69 | $Mn_{1.00}Pd_{0.94}Pt_{4.11}P_{0.80}$ |
| | | | | | ***Average*** | $\mathbf{Mn_{1.00}Pd_{0.92}Pt_{3.96}P_{0.82}}$ |

**Figure S2.** Comparison of magnetic susceptibility vs T for each compound at 1000 Oe

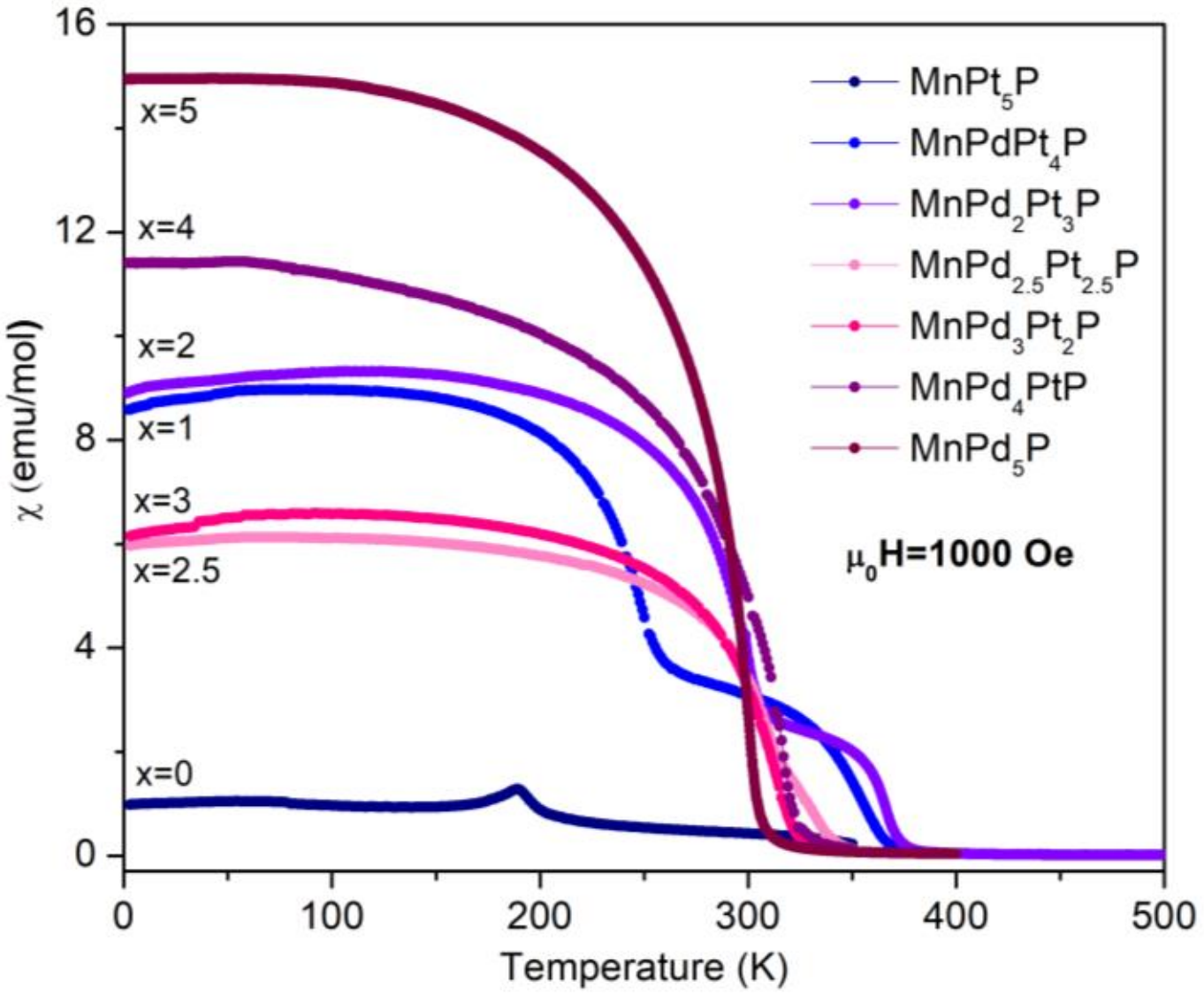

**Figure S3.** Summary of magnetic parameters derived from Curie-Weiss fitting

| | $MnPd_5P$ | $MnPd_4PtP$ | $MnPd_3Pt_2P$ | $MnPd_{2.5}Pt_{2.5}P$ | $MnPd_2Pt_3P$ | $MnPdPt_4P$ |
|---|---|---|---|---|---|---|
| $\theta_{CW}$ (K) | 304.4(1) | 317.8(3) | 318.0(1) | 334.1(2) | 369.9(1) | 366.2(3) |
| C | 3.13(1) | 4.14(6) | 2.45(1) | 2.09(4) | 1.09(1) | 1.16(1) |
| $\mu_{eff}$ ($\mu_B$) | 5.00 | 5.75 | 4.43 | 4.09 | 2.95 | 3.04 |
| $R^2$ | 0.999 | 0.992 | 0.999 | 0.989 | 0.998 | 0.989 |

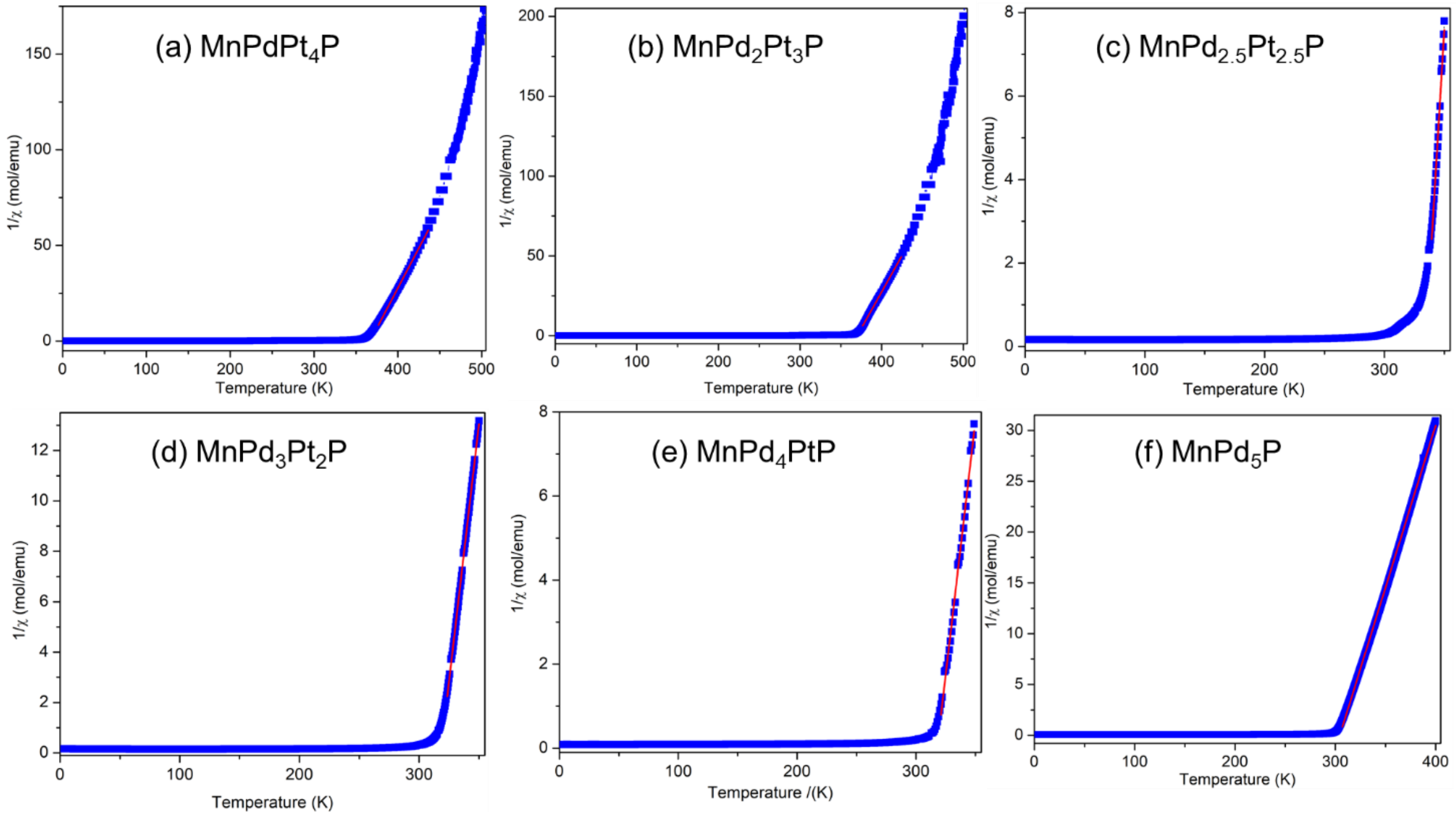